\documentclass[prd,twocolumn]{revtex4}

\usepackage{graphicx}
\usepackage{amsmath,amssymb,amsfonts}
\newcommand{\be}{\begin{equation}}
\newcommand{\ee}{\end{equation}}
\newcommand{\ba}{\begin{eqnarray}}
\newcommand{\ea}{\end{eqnarray}}
\newcommand{\ban}{\begin{eqnarray*}}
\newcommand{\ean}{\end{eqnarray*}}
\newcommand \nn {\nonumber}

\begin{document}
\author{Berndt M\"uller}
\affiliation{Department of Physics, Duke University, Durham, NC 27708-0305, USA}
\author{Andreas Sch\"afer}
\affiliation{Institut f\"ur Theoretische Physik, Universit\"at Regensburg, D-Regensburg, Germany}

\title{The Chiral Magnetic Effect and an experimental
  bound on the late time magnetic field strength}

\begin{abstract}
We first compare different approaches to estimates of the magnitude of the chiral magnetic effect in relativistic heavy ion collisions and show that their main difference lies in the assumptions on the length of persistence of the magnetic field generated by the colliding nuclei. We then analyze recent measurements of the global polarization of $\Lambda$ and $\bar \Lambda$ hyperons in terms of the bounds they set on the magnitude of the late time magnetic field.
\end{abstract}

\maketitle

\section{Introduction}

A few years ago the claim that experimental data from the STAR collaboration \cite{Abelev:2009ac,Abelev:2009ad} could indicate a sizeable Chiral Magnetic Effect (CME)
\cite{Fukushima:2008xe, Kharzeev:1998kz} in peripheral heavy ion collisions initated extensive theoretical and experimental research. Recently, new analyses of data from heavy ion collisions at the Relativistic Heavy Ion Collider (RHIC) and the CERN Large Hadron Collider (LHC) by several experiments (STAR \cite{STAR-new}, ALICE \cite{Acharya:2017fau}, CMS \cite{Sirunyan:2017quh,Tu:2018ham}) have been presented, which conclude that only a small fraction of the observed charge imbalance fluctuations can be attributed to the CME.  On the experimental side, a dedicated run of the $^{96}\mathrm{Zr}+^{96}\mathrm{Zr}$ and $^{96}\mathrm{Ru}+^{96}\mathrm{Ru}$ isobar systems at RHIC this year \cite{Wen:2018boz} has provided new data which are expected to yield unambiguous evidence for (or against) the presence of a CME contribution at the few percent level.

Our contribution is intended to add a new and supplementary angle to this discussion. In doing so we will not try to review any significant part of the very extensive literature on this topic. Instead, based on a recent analysis  of topological fluctuations in the glasma \cite{Lappi:2017skr} we will reiterate the observation that although the underlying physics is quite complicated, different approaches give similar results leading to the conclusion that whether or not phenomena related to the CME can be large enough to be observable in relativistic heavy ion collisions depends crucially on the longevity of the magnetic field. As originally proposed by Tuchin \cite{Tuchin:2015oka}, and explored further by G\"ursoy {\em et al.} \cite{Gursoy:2014aka}, the magnetic field may be partially ``frozen in'' by the large electric
conductivity and the non-linear properties of the quark gluon plasma. Such a delayed decay could result in a significant magnetic field strength at mid-rapidity and late times.

Below we point out that new limits on the difference in observed global $\Lambda$ and $\bar\Lambda$ polarization transverse to the reaction plane in peripheral heavy ion collisions, a phenomenon that is at the focus of intense investigation in connection with the study of the quark gluon plasma vorticity
\cite{STAR:2017ckg,BedangadasMohantyfortheALICE:2017xgh,Adam:2018ivw}, provides a relevant limit on the magnetic field strength at late times and thus a direct test for, e.g., the prediction made in \cite{Gursoy:2014aka}.

Consequently, our short note has two parts. In Section II we will compare the outcome of three principally different estimates for the topologic charge fluctuations generated in a heavy ion collision, namely one based on glasma phenomenology \cite{Lappi:2017skr}, one based on heuristic arguments made by us \cite{Muller:2010jd}, and one that is based on quantitative
simulation of anomalous hydrodynamics \cite{Hirono:2014oda}.

In Section III we will derive a bound on the late time magnetic field strength from the present non-observation of a global polarization difference for $\Lambda$ and $\bar\Lambda$ produced in peripheral heavy ion collisions.

Section IV summarizes our conclusions and provides an outlook for the potential impact of future data.

\section{Quantitative estimates of the CME at RHIC energies} 

We will split our discussion into two parts. First we compare different estimates for the average density of the topological winding number fluctuation $\overline {Q_5^2}/V^2$, which turn our to give very similar results. Then we will compare the resulting predictions for the charge asymmetry with respect to the collision plane.

\subsection{Estimates of the topological charge density}

\subsubsection{Glasma approach}

Let us start with the recent results of Lappi amd Schlichting \cite{Lappi:2017skr} (referrred to as LS) on topological charge ($Q_5$) fluctuations in high-energy heavy ion collisions. Starting from Eq.~(5.3) in LS one can estimate the mean square value of $Q_5$ in a space-time rapidity interval $\Delta\eta$ as follows:
\ba
\overline{(Q_5)^2} &=& \bigg\langle \int \frac{dN_5}{d^2{\bf x} d\eta} d^2{\bf x}~\int \frac{dN_5}{d^2{\bf y} d\eta} d^2{\bf y}\bigg\rangle (\Delta\eta)^2
\\ 
&\approx& \frac{3\alpha_s^2N_f^2}{8\pi^2(N_c^2-1)} ~\pi(\rho\tau^2\Delta\eta)^2 \int d^2x ~\varepsilon(x,\tau)^2 , 
\nn
\label{eq:Q5sq}
\ea
where $\rho$ is the transverse correlation length of topological charge density given by
\be
\rho^2 = \frac{1}{\pi} \int d^2z \left( \frac{1 - e^{-a Q_\textrm{s}^2\vert{\bf z}\vert^2}}{a Q_\textrm{s}^2\vert{\bf z}\vert^2} \right)^4 \approx 0.99~Q_\textrm{s}^{-2} 
\ee
with $a = N_c/4C_\textrm{F} = 9/16$ and the saturation scale $Q_s$. $\eta$ is the space-time rapidity, and the metric is $g_{\mu\nu}=(1,-\tau^2,-1,-1)$ in the $(\tau,\eta,X_{\perp})$ coordinate system, see the section after Eq. (9) in \cite{Lappi:2006hq}. The expression \eqref{eq:Q5sq} is valid for $\tau \leq 1/Q_s$. In the following, we shall assume
\be
\int d^2x ~\varepsilon(x,\tau)^2 \approx \varepsilon(\tau)^2 S_\perp ,
\ee
where $S_{\perp}$ denotes the transverse overlap area of the colliding nuclei. We further identify $\tau \Delta\eta ~S_{\perp}$ with the total volume $V$.  Inserting this into \eqref{eq:Q5sq} one obtains
\be
\overline{(Q_5)^2} \approx \frac{3\alpha_s^2N_f^2}{8\pi^2(N_c^2-1)} ~ V^2 \tau^2 \varepsilon(\tau)^2 ~\frac{\pi\rho^2}{S_\perp} .
\ee

The product $\tau\varepsilon(\tau)$ is approximately time independent in the early, pre-hydrodynamic phase of the reaction. We therefore estimate its value from that prevailing at the onset of the hydrodynamic phase of the quark-gluon plasma. Expressing this ``fluidization'' time $\tau_0$ in terms of the initial temperature $T_0$ as $\tau_0 \approx 0.7/T_0$ and using $\varepsilon \approx \frac{45}{30}\pi^2 T^4$ for a quark-gluon plasma with three light flavors, we obtain
\ba
\overline{(Q_5)^2} & \approx & \frac{3\alpha_s^2N_f^2}{8\pi^4(N_c^2-1)} ~ V^2 (\pi T_0)^6 ~\frac{\pi\rho^2}{S_\perp}
\nn \\
& \approx & 0.0005 ~{\rm GeV}^6~V^2~\frac{\pi\rho^2}{S_\perp} ,
\label{eq:Q5-T0}
\ea
where we set $\pi T_0 \approx 1$ GeV.

\subsubsection{Heuristic approach}

In \cite{Muller:2010jd} we had estimated the volume density of winding number transitions geometrically as
\be
|Q_5|/V ~=~ (\overline\rho_{\rm top})^{-3}
\ee
with a phenomenologically motivated average size of cells with correlated topological charge density in the quark-gluon plasma phase $\overline \rho_{\rm top}\approx 0.5~{\rm fm}$. The total $Q_5$ charge is generated by winding number fluctuations which increase as the square root of the number of independent domains, i.e. $\sqrt{V/\rho_{\rm top}^3}$ in LS this number is
$V^2\frac{\pi\rho^2}{S_{\perp}}$. Thus we have to compare
\be
(\rho_{\rm top})^{-6}  ~=~ (0.4~\textrm{GeV})^6 = 0.004~\textrm{GeV}^6
\label{eq:Q5-MS}
\ee
with the number given in (\ref{eq:Q5-T0}). Since $\rho_{\rm top}$, which is not known precisely, enters to the sixth power, the factor 8 difference between the results \eqref{eq:Q5-T0} and \eqref{eq:Q5-MS} hardly constitutes a serious discrepancy.

\subsubsection{Anomalous hydrodynamics}

In \cite{Hirono:2014oda} the topological charge fluctuations were linked to the expectation value of the QCD analogues of the electric and magnetic field strengths in a manner that corresponds to 
\be
\overline{(Q_5)^2} \approx \frac{1}{(4\pi)^4} ~ V^2 \tau_0^2 (gE^agB^a)^2 ~\frac{\pi\rho^2}{S_\perp} .
\ee
With $\vert gE^a \vert \approx \vert gB^a \vert \approx Q_\textrm{s}^2$ this gives 
\be
\overline{(Q_5)^2} \approx \frac{1}{(4\pi)^4} ~ V^2 Q_\textrm{s}^6 ~\frac{\pi\rho^2}{S_\perp} .
\label{eq:Q5-Qs}
\ee
The expression to be compared with that in \eqref{eq:Q5-T0} is:
\ba
\frac{1}{(4\pi)^4} ~Q_\textrm{s}^6 &\approx& \frac{1}{(4\pi)^4} ~(2~\textrm{GeV}^2)^3 \approx 0.0003~\textrm{GeV}^6 .
\ea
in far better agreement with \eqref{eq:Q5-T0} than could have been expected based on the large uncertainties entering any such estimate.

\subsection{Estimate of charge asymmetry fluctuations}

To set the stage for the continuation of our discussion we remind the reader that in \cite{Muller:2010jd} we obtained as relative charge asymmetry
\be
\Delta^{\pm} \approx 3.4\times 10^{-7} ~\frac{b}{R}
\label{eq:Delta-pm}
\ee
with impact parameter $b$ and nuclear radius $R$, which is three orders of magnitude too small to explain the STAR data \cite{Abelev:2009ac,Abelev:2009ad}. Our conclusion was that the CME could at most contribute on the order of a percent of the observed asymmetry. We note that this estimate contains a correction factor $1/f(B)\approx 1/10$ which was introduced in \cite{Muller:2010jd} in the attempt to take nonlinear electrodynamic effects into account. The estimate of this factor was on somewhat shaky grounds at the time of our original publication, but the assumed value $f\approx 10$ was later confirmed by a dedicated lattice QCD simulation \cite{Bali:2014vja}. (While the argument leading to this suppression factor are quite different in both cases they both lead to the suppression of the same mixed correlation $F_{\mu}\tilde F^{mu\nu} ~G_{lambda\nu}^a G^{a\lambda\nu}$.) This factor compensates the larger estimate of (7) compared to the other two, such that they all nearly coincide. However, this degree of agreement must be viewed as purely accidental.

The conclusion \eqref{eq:Delta-pm} contrasts sharply with that of \cite{Hirono:2014oda} which was that ``the experimental observations are consistent with the presence of the effect.'' Some part of this
discrepancy can be caused by the very different treatment of the subsequent dynamics of the produced fireball. In \cite{Muller:2010jd} this involved the introduction of a chiral chemical potential combined with a straightforward application of thermodynamics while HHK performed a numerical simulation of anomalous hydrodynamics. By far the largest difference between the two estimates was the assumed time evolution of the average magnetic field strength in peripheral heavy ion collisions.

MS estimated the average magnetic field factor by (see eq.~(26)):
\be
\tau_B e {\bar B} \approx \frac{2.3 Z\alpha\, b}{R^2} \approx 0.04~\textrm{GeV} .
\ee
On the other hand, HHK have (using their value $\tau_B = 3$ fm):
\be
\tau_B e {\bar B} = e B_0 \frac{b}{2R} \int d\tau\, e^{-\tau/\tau_B} = \tau_B e B_0 \frac{b}{2R}
\approx 4~\textrm{GeV} .
\ee
The difference of a factor 100 comes from the much longer persistence of the magnetic field assumed by HHK and a somewhat larger peak field. (In MS the estimate for the maximum magnetic field is $\tau_B \approx 2R/\gamma_{\rm cm} \approx 0.15$ fm at RHIC.) 

The lifetime of the magnetic field $\tau_B$ enters also the factor $C_{\rm em}$, which converts the magnetic effect into an electric charge fluctuation. The strength of this conversion factor again carries a factor $\tau_B$, because it only operates as long as the magnetic field is present. MS estimates the conversion factor as
\be 
C_{\rm em} = \frac{8\alpha \tau_B v}{\pi T_f^3 \sqrt{N}} \approx \frac{2~\textrm{GeV}^{-4}}{\sqrt{N}} .
\ee
Using the assumptions of HHK, this number would be a factor 20 larger (because their value of $\tau_B$ is a factor 20 larger than that of MS. Since the HHK calculation is based on an entirely different formalism it is not clear how large precisely the effective statistical factor is in their approach but our estimate should give the correct magnitude.

We conclude that the difference between the various predictions of the magnitude of the chiral magnetic effect is mainly in how long the magnetic field is assumed to persist. HHK's estimate is based on the argument that the electrically conducting quark-gluon plasma ``freezes'' the magnetic field such that it decays slowly, see Tuchin \cite{Tuchin:2015oka} and also G\"ursoy, Kharzeev and Rajagopal \cite{Gursoy:2014aka}. The reason why the estimate by G\"ursoy et al. is significantly larger than that by Tuchin is probably due to the fact that Tuchin does not assume that a conducting medium is present from the very beginning, thereby reducing the magnitude of the initial current imprinted on the medium.

Altogether, we find that the estimates of LS, MS and HHK on the topological charge density agree better than could reasonably be expected, but MS and HHK differ markedly in their predicted charge separation signal. This difference, however, is tied to the assumed evolution of the magnetic field strength. For MS the time-integrated magnetic field strength is 100 times smaller and the electric conversion factor is smaller by an order of magnitude. Together this explains the large difference between the  HHK value for the electric charge asymmetry of order $2\times 10^{-4}$ and the MS estimate of $3.5\times 10^{-7}$.

As it is difficult to perform an {\em ab-initio} calculation of the time evolution of the magnetic field induced by the colliding nuclei and its delayed decay in the electrically conducting quark-gluon plasma, the search for direct experimental evidence for a long-lasting remnant of the magnetic field is well motivated.  In the following section we explore such a signal.

\section{Bounds on the late-time magnetic field}

The recent discovery of a substantial global transverse polarization of $\Lambda$ and $\bar\Lambda$ produced in peripheral heavy ion collisions has led to the conclusion, that the quark gluon plasma is the ``most vortical fluid'' ever observed \cite{STAR:2017ckg}. For this interpretation it is of key importance that the polarization of $\Lambda$ and $\bar\Lambda$ is the same within experimental errors, ruling out the presence of any sizeable magnetic field effect during the emission of the hyperons. This observation can be turned into an upper bound for the magnetic field strength at the time of hadronization of the quark-gluon plasma.

Because the $\Lambda$ polarization is mainly determined by the polarization of its valence strange quark one can imagine two origins of the polarization. The $s$ and $\bar s$ quarks can get produced and polarized in the quark-gluon plasma at early times, wenn the $B$-field has maximum strength and later coalesce to form polarized $\Lambda$ and $\bar\Lambda$ hyperons, or the hyperons can become statistically polarized at the time of hadronization. Here we will only discuss the latter mechanism such that our results should be seen as a conservative upper bound on the allowed late-time contribution and thus the late-time magnetic field strength.

The difference in the global polarization of $\Lambda$ hyperons and their antiparticles is defined as \cite{STAR:2017ckg}
\be
\Delta{\cal P} = {\cal P}_\Lambda - {\cal P}_{\bar\Lambda} ,
\ee
where the individual polarizations are given by the relative differences in the emission yields of hyperons polarized perpendicular to the collision plane:
\be
{\cal P}_i = \frac{N_i^\uparrow - N_i^\downarrow}{N_i^\uparrow + N_i^\downarrow} ,
\ee
where $i$ denotes $\Lambda$ or $\bar\Lambda$ and $\uparrow, \downarrow$ denotes particles with spin orientation parallel or antiparallel to the global angular momentum vector of the colliding nuclear system. The relative yield of thermally emitted hyperons with spin $\vec s$ is given by \cite{Becattini:2016gvu}
\be
N_i({\vec s}) \propto \exp\left({\vec s}\cdot{\vec\omega}/T_s + 2\mu_i{\vec s}\cdot{\vec B}/T_s\right) ,
\ee
where $\vec\omega$ denotes the vorticity of the matter, $\mu_i$ the magnetic moment of the (anti-)hyperon, $\vec B$ the magnetic field present at the moment of particle emission, and $T_s$ is the temperature of the emitting source. Since the exponent is small compared to unity, it is sufficient to keep the linear term in the expansion of the exponentials, resulting in 
\ba
{\cal P}_\Lambda \approx \frac{\omega}{2T_s} + \frac{\mu_\Lambda B}{T_s} ,
\nn \\
{\cal P}_{\bar\Lambda} \approx \frac{\omega}{2T_s} - \frac{\mu_\Lambda B}{T_s} ,
\ea
where we used the fact that $\mu_{\bar\Lambda} = - \mu_\Lambda$ and $\omega$ and $B$ denote the magnitude of the vorticity and magnetic field, respectively. Thus
\be
\Delta{\cal P} = \frac{2\mu_\Lambda B}{T_s} .
\ee

The most precise limit on the late-time magnetic field comes from a recent measurement of the global polarization of $\Lambda$ and $\overline\Lambda$ hyperons in Au+Au collisions at $\sqrt{s_{\rm NN}} = 200$ GeV \cite{Adam:2018ivw}:
\ba
{\cal P}_\Lambda = (0.277 \pm 0.040) \times 10^{-2} ,
\nn \\
{\cal P}_{\bar\Lambda} = (0.240 \pm 0.045) \times 10^{-2} ,
\ea
corresponding to 
\be
\Delta{\cal P} = (0.37 \pm 0.60) \times 10^{-3} .
\ee
At the one-standard deviation limit, thus $-\Delta{\cal P} < 2.3 \times10^{-4}$. With $\mu_\Lambda = - 0.613 \mu_N = - 1.93 \times 10^{-14}$ MeV/T and $T_s \approx 150$ MeV, we obtain
\be
|B| = \frac{T_s~|\Delta{\cal P}|}{2|\mu_\Lambda|} < 8.9 \times 10^{11}~\mathrm{T} ,
\ee
where we took into account that the collision induced magnetic field must point in the same direction as the angular momentum of the collision system and hence the vorticity vector, implying $\Delta{\cal P} < 0$. This corresponds to a limit of $|eB| < 1.35 \times 10^{-3}~\textrm{fm}^{-2} \approx 2.7 \times 10^{-3}~m_\pi^2$ (in natural units $\hbar=c=1$). Comparing this value with Fig.~2 in \cite{Gursoy:2014aka} and Fig.~12 in \cite{Inghirami:2016iru} shows that this bound is already significant and rules out all models of plasma induced magnetic field persistence considered in Inghirami {\em et al.} \cite{Inghirami:2016iru}. More precise data could further sharpen this constraint. 

It is useful to consider the implications of the bound on the late-time magnetic field on the time integrated magnetic field $\tau_B e\bar{B}$ which governs the observable size of the chiral magnetic effect. We assume an exponential decay of the magnetic field $B(t) = B_0 \exp(-t/\tau_B)$ with $eB_0 \approx 0.5 m_\pi^2$ \cite{Inghirami:2016iru}.  The lifetime of the quark-gluon plasma in a 200~GeV Au+Au collision is $t_s \approx 5~\textrm{fm}/c$ \cite{Song:2007ux}. Using the just derived limit $eB(t_s) < 0.0027~m_\pi^2$ on the magnetic field at hadronization we then obtain
\be 
\tau_B = t_s \left(B_0/B(t_s)\right)^{-1} \approx 1~\textrm{fm}/c 
\ee
resulting in the estimate
\ba 
\tau_B e\bar{B} & \equiv & \int_0^{t_s} eB(t)~dt =\tau_B eB_0 
\nn \\
&<& 0.25~\textrm{fm}^{-1} \approx 50~\text{MeV} .
\ea
This bound is only modestly larger than the time integrated value of the vacuum magnetic field (see Fig.~11 in Asakawa {\em et al.} \cite{Asakawa:2010bu}). We conclude that the current limit already precludes a significant enhancement of CME phenomena due to plasma effects on the lifetime of the magnetic field.

\section{Summary}

In summary, we have argued that different approaches to calculating the initial axial charge density fluctuations in the quark-gluon plasma created in heavy ion collisions yield estimates of similar magnitude. We then traced the widely different published estimates of the contribution of the chiral magnetic effect to various experimental observables associated with the CME to disparities in the assumption of persistent magnetic fields that survive until hadronization. Finally, we pointed out that recent data on the global polarization of hyperons in heavy ion collisions yield an upper bound on the late-time magnetic field that is severely constraining model predictions of the chiral magnetic effect.

{\em Acknowledgments:} We thank M.~Lisa for helpful advice concerning the uncertainties of the STAR data on the global hyperon polarization and L.~Yaffe for comments on an earlier version of the manuscript. This work was supported by DOE grant DE-FG02-05ER41367 and BMBF grant O5P15WRCCA. One of us (AS) thanks the Nuclear and Particle Physics Directorate of Brookhaven National Laboratory for its hospitality.

\end{document}